\documentclass[twocolumn,showpacs,preprintnumbers,amsmath,amssymb]{revtex4-1}

\usepackage{graphicx}
\usepackage{dcolumn}
\usepackage{bm}

\usepackage{epsfig}

\begin{document}

\title{THz emission from a stacked coherent flux-flow oscillator: non-local radiative boundary conditions and the role of geometrical resonances.}

\author{V. M. Krasnov}
\affiliation{
Department of Physics, Stockholm University, AlbaNova University Center, SE-10691 Stockholm, Sweden
}

\date{\today}

\begin{abstract}
I derive simple non-local dynamic boundary conditions, suitable
for modelling of radiation emission from stacked Josephson
junctions, and employ them for analysis of flux-flow emission from
intrinsic Josephson junctions in high-$T_c$ superconductors. It is
shown that due to the lack of Lorenz contraction of fluxons in
stacked junctions, high quality geometrical resonances are
prerequisite for high power emission from the stack. This leads to
a dual role of the radiative impedance: on the one hand, small
impedance increases the efficiency of emission from the stack, on
the other hand, enhanced radiative losses reduce the quality
factor of geometrical resonances, which may decrease the total
emission power. Therefore, the optimal conditions for the coherent
flux-flow oscillator are achieved when radiative losses are
comparable to resistive losses inside the stack.

\end{abstract}

\maketitle


Creation of a compact, high power THz source remains a difficult
technological challenge, colloquially known as ``the THz gap"
\cite{THzGap}. Flux-flow oscillators (FFO's), based on regular
motion of quantized vortices (fluxons) in Josephson junctions, can
generate tunable THz radiation with a remarkable linewidth
$10^{-12}$, albeit with a low emission power $<1 \mu W$
\cite{Koshelets}. The power could be greatly enhanced by coherent
phase-locking of several coupled FFO's \cite{Barbara,SakUst}. The
required coupling is strongest in stacked, atomic scale intrinsic
Josephson junctions (IJJs), naturally formed in layered high
temperature superconductor
$\text{Bi}_{2}\text{Sr}_{2}\text{Ca}\text{Cu}_{2}\text{O}_{8+x}$
(Bi-2212). IJJs allow simple integration of a large number of
almost identical stacked Josephson junctions. Furthermore, a large
superconducting energy gap  in Bi-2212 \cite{SecondOrder}
facilitates operation in the important THz frequency range (up to
$\sim$ 10THz). Therefore, IJJs are actively studied both
theoretically
\cite{FFlowSimul,FFlowMachida,Bul2007,Cascade,Hu,Savelev} and
experimentally
\cite{Cherenkov,Batov,Bae2007,Ozyuzer,FiskeInd,SvenFiske} as
possible candidates for realization of a coherent THz oscillator.

Proper radiative boundary conditions are essential for analysis of
the coherent FFO. For a single junction this can be done by
introducing an appropriate radiation impedance $Z$ \cite{Bul2006},
connecting {\it local} ac-components of electric and magnetic
fields at the junction edges.
\begin{equation}\label{Z}
Z=Z(\omega)=E_{ac}/H_{ac}.
\end{equation}
The emission power is $P_{rad}=V_{ac}^2/R_Z$, where
$V_{ac}=t_0E_{ac}$ is the ac-voltage, $t_0$ is the junction
barrier thickness and
\begin{equation}\label{RZ}
R_Z=(t_0/w)Z
\end{equation}
is the radiative resistance of a single junction, $w$ is the width
of the junction.
However, the stacked FFO can not be described by a single $Z$
because the effective radiative resistance depends not only on the
geometry of the stack, but also in a crucial way on the collective
fluxon configuration. For example, motion of the triangular fluxon
lattice corresponds to $Z\rightarrow\infty$ because it results in
out-of-phase oscillations 
in neighbor junctions $E_i=-E_{i+1}$,
leading to destructive interference and negligible emission. On
the other hand, motion of the rectangular lattice corresponds to
small $Z$ because it results in in-phase oscillation
$E_i=E_{i+1}$, leading to constructive interference and coherent
enhancement of the emission power $\propto N^2$, where $N$ is the
number of junctions in the stack. Such behavior is caused by the
essentially {\it non-local} nature of magnetic induction
\cite{Fluxon}, which is the consequence of the inductive coupling
of junctions in the stack. Therefore, the relation between local
$E_{ac}$ and non-local $H_{ac}$
can not be described by Eq. (\ref{Z}).

In this letter I derive simple non-local dynamic radiative
boundary conditions for stacked Josephson junctions and employ
them for analysis of flux-flow emission from IJJ-based stacked
FFO. It is shown that due to the lack of Lorenz contraction of
fluxons in stacked junctions \cite{Fluxon}, high quality $Q \gg 1$
geometrical resonances are prerequisite for high power emission
from the stack. This leads to a dual role of the radiative
impedance: on the one hand, small $Z$ increases the emission
efficiency, but on the other hand, enhanced radiative losses
reduce $Q$ of geometrical resonances, which leads to reduction of
the emission power. The maximum power is achieved when radiative
losses are equal to internal losses in the stack.

Neumann (static) boundary conditions are most widely used for
numerical modelling of Josephson junctions:
\begin{equation}\label{Neumann} \frac{\partial \varphi}{\partial
x} (x=0,L) = B \frac{2\pi \lambda_J \Lambda^*}{\Phi_0}.
\end{equation}
Here $\varphi$ is the Josephson phase difference, $x=0,L$ are
coordinates of junction edges, $B$ is the applied dc-magnetic
field, $\lambda_J$ is the Josephson penetration depth and
$\Lambda^*$ is the effective magnetic thickness of the junction
\cite{Fluxon}. As emphasized in Ref. \cite{Bul2006}, Neumann
boundary conditions are non-radiative because they assume
$H_{ac}(0,L)=0$.

Radiative (dynamic) boundary conditions should account for the
finite $H_{ac}$. For a single junction they can be easily written
with the help of Eq. (\ref{Z}) \cite{Hu,Bul2006}:

\begin{equation}\label{RadBounCond}
\frac{\partial \varphi}{\partial x} (x=0,L) = \left( B \pm \mu_0
\frac{E_{ac}(0,L)}{Z} \right) \frac{2\pi \lambda_J
\Lambda^*}{\Phi_0}.
\end{equation}
Here plus and minus signs correspond to $x=0$ and $L$,
respectively, because the direction of emission is opposite at
opposite edges of the junction.

However, Eq.(\ref{RadBounCond}) is not directly applicable for
stacked junctions due to the principle difference in
electrodynamics, which is local in single, and non-local in
stacked junctions. The electric field $E_{ac}$ is local in each
junction because the Debye screening length is always smaller than
the electrode thickness $d$, even for IJJs.
However, the magnetic field in inductively coupled stacked
junctions is non-local and is created and shared cooperatively by
the whole stack \cite{Fluxon}. The non-locality of $H_{ac}$ is
particularly dramatic for IJJs, for which the London penetration
depth $\lambda_{ab} \simeq 200$ nm is much larger than $d < 1$ nm.

To derive the proper non-local radiative boundary conditions for
stacked Josephson junctions, lets note that the local relation
Eq.(\ref{Z}) is valid outside the stack, however, here the
electric field is the result of interference of electric fields
from all junctions:
\begin{equation}\label{Enonlocal}
E_{ac}=\sum_{i=1}^N E_i.
\end{equation}

The net emission power from one edge of the stack is
\begin{equation}\label{EmPower}
P_{rad}=wt_0 H_{ac}\sum_{i=1}^N E_i = wt_0E_{ac}^2/Z.
\end{equation}
Radiative losses are associated with additional currents flowing
through edges of the stack:
\begin{equation}\label{Irad}
\Delta I_{rad} =wE_{ac}/Z
\end{equation}
Those displacement-like currents should be added in the numerical
scheme for the sake of energy conservation. Equations
(\ref{RadBounCond},\ref{Enonlocal},\ref{Irad}) together form the
final non-local dynamic boundary conditions for stacked Josephson
junctions. For the out-of-phase state, $E_i = -E_{i+1}$, $E_{ac} =
0$, they reduce to the non-radiative Neumann condition
Eq.(\ref{Neumann}). For the in-phase state, $E_i = E_{i+1}$,
$E_{ac} = N E_i$, they lead to coherent power amplification
$\propto N^2$, Eq.(\ref{EmPower}).

To calculate flux-flow emission, the non-local boundary conditions
were implemented into the coupled sine-Gordon equation
\cite{SakUst}, which describes electrodynamics of inductively
coupled stacked Josephson junctions (see Refs.\cite{Modes,Fluxon}
for details of the formalism and the Supplementary \cite{Supplem}
for technical details). Simulations were made for $N=10$ identical
IJJs with $L=4 \lambda_J \simeq 2.8 ~\mu$m, $w=1.4 ~\mu$m, the
stacking periodicity $s=1.55$ nm, the electrode thickness $d=0.6$
nm, the ratio of the barrier thickness to the relative dielectric
constant $t_0/\epsilon_r \simeq 0.11$ nm,
$\lambda_{ab} = 200$ nm, and the critical current density
$J_{c0}=1050$ A/cm$^2$. Those parameters correspond to $\lambda_J
\simeq 0.7 \mu$m, the Josephson plasma frequency $\omega_P\simeq
6.4 \times 10^{11}$ 1/s, and the slowest velocity of light $c_N
\simeq 3.2 \times 10^5$ m/s, typical for small Bi-2212 mesa
structures \cite{Katterwe,SvenFiske}. The damping parameter
$\alpha =(\omega_p R C)^{-1} =0.05$, corresponds to junction
resistance $R\simeq 100 ~\Omega$ and the $c-$axis resistivity
$\rho_c \simeq 25 ~\Omega$cm, equal to large bias tunnel
resistivity of IJJs \cite{Doping}. However, the quasiparticle (QP)
resistivity at small bias and low $T$ can be up to two orders of
magnitude larger \cite{Katterwe2008}. Therefore, $\alpha =0.05$
represents the highest limit of QP damping in IJJs. The correct
$\alpha$ is important for numerical modelling because it
determines $Q \propto 1/\alpha$ in the absence of radiative
losses. Simulations were also performed for smaller $\alpha$ down
to $0.005$, which show qualitatively similar behavior
\cite{SvenFiske}, but require substantially longer integration
times. However, for larger $\alpha =0.1$ the number and amplitudes
of accessible geometrical resonances were significantly reduced.

\begin{figure}[t]
\begin{center}
\includegraphics[width=0.9\linewidth]{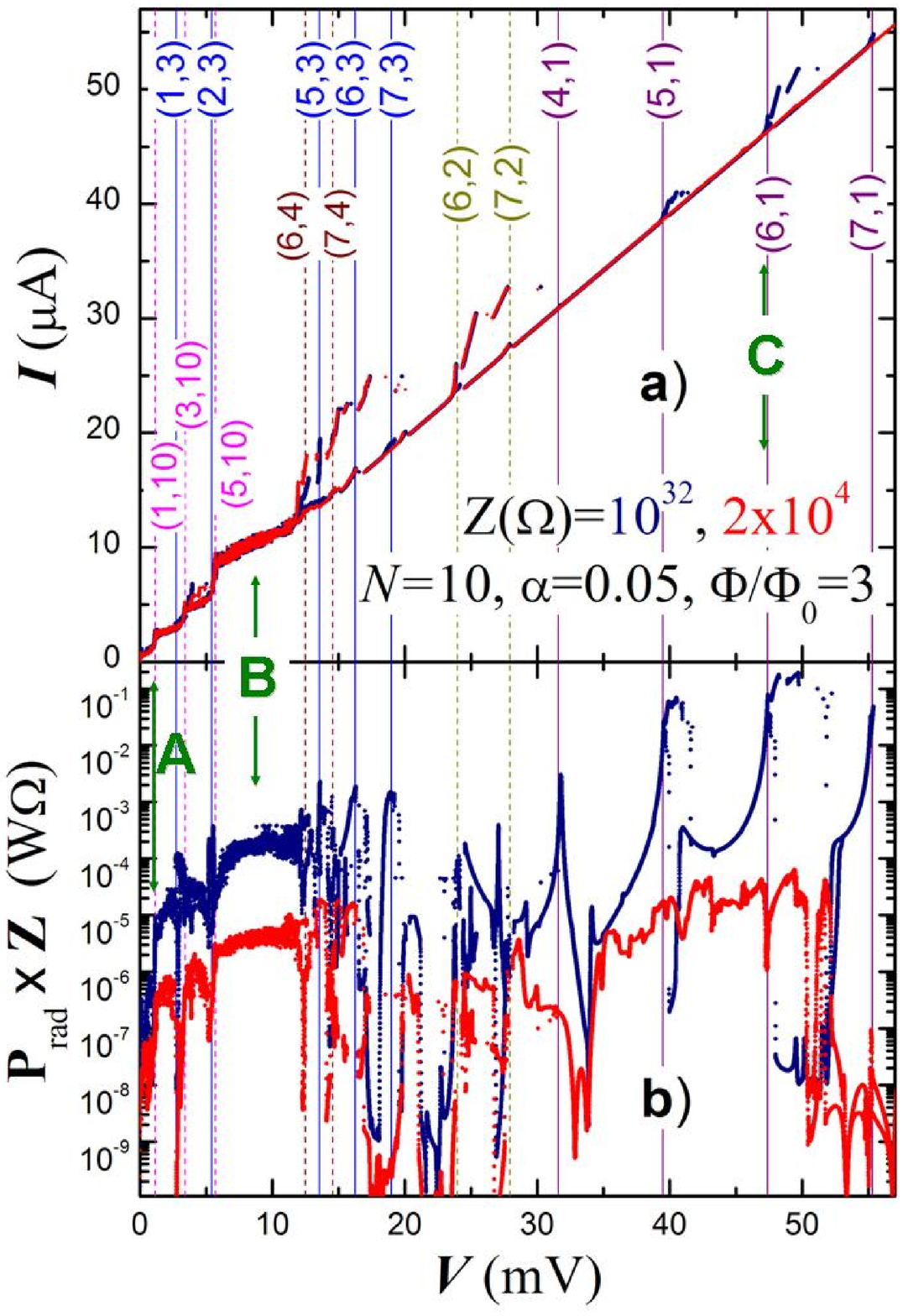}
\end{center} \label{Fig1}
\caption{(color online). Simulated a) $I-V$ characteristics and b)
normalized radiative powers for negligible $Z=10^{32}~\Omega$
(dark blue) and moderate $Z=2 \times 10^4~\Omega$ (red) emission.
A variety of Fiske steps is seen. Mode numbers $(m,n)$ for the
most prominent steps are indicated. Note that non-radiating
even-$n$ steps (dashed vertical lines) are not affected by $Z$,
while radiating odd-$n$ steps (solid lines) vanish for small $Z$.}
\end{figure}

Fig. 1 a) shows simulated $I-V$ characteristics for two impedance
values, corresponding to negligible $Z=10^{32} ~\Omega$ (dark
blue) and moderate $Z=2\times 10^{4} ~\Omega$ (red curve)
radiation losses, at $\Phi/\Phi_0=3$, where $\Phi=BLs$ is the flux
per junction. Data for $Z=10^{32}~\Omega$ are similar to
simulations with non-radiative Neumann boundary conditions
\cite{FFlowSimul,FFlowMachida,SvenFiske}. A large variety of Fiske
steps is seen.

Geometrical (Fiske) resonances are caused by resonant excitation
of phonon-like collective fluxon lattice vibrations
\cite{SvenFiske} in the flux-flow state. In stacked junctions such
fluxon phonons are two-dimensional and are characterized by two
wave numbers $k_m=\pi m /L ~(m=1,2,3,...)$ in plane, and $q_n=\pi
n/Ns ~(n=1,2,...N)$ in the $c$-axis direction \cite{KleinerModes}.
Unlike Josephson plasma waves, fluxon phonons have a linear
dispersion relation at low frequencies. Fluxon phonons with the
wave number $q_n$ propagate with the in-plane velocity $c_n$
\cite{KleinerModes}. The slowest, $c_N$, corresponds to the
out-of-phase, $q_N=\pi/s$, and the fastest, $c_1$, to the
in-phase, $q_1=\pi/Ns$, mode. Fiske steps occur when the
ac-Josephson frequency coincides with the frequency of one of the
cavity modes $(m,n)$ \cite{SvenFiske}:
\begin{equation}\label{Fiske}
V_{m,n}=\Phi_0 m c_n/2L ,~(m=1,2,3..., ~n=1,2...N).
\end{equation}

In Fig. 1, voltages of the most prominent phase-locked Fiske steps
$NV_{m,n}$ are marked by vertical lines together with mode numbers
$(m,n)$. For the non-radiative case $Z=10^{32}~\Omega$, Fiske
steps with both even and odd $n$ are equally represented. However,
for $Z=2\times 10^{4}~\Omega$ we observe a clear difference
between even and odd $n$ resonances: even-$n$ steps, e.g. (1,10),
(3,10), (5,10), (6,4), (7,4), (6,2) and (7,2) are practically the
same as for $Z=10^{32} ~\Omega$, but all odd-$n$ steps are
strongly reduced. For example, steps (5,3), (5,1) and (6,1) are
well developed for $Z=10^{32} ~\Omega$, but not present for
$Z=2\times 10^{4}~\Omega$.

The observed difference between even and odd-$n$ steps for
$Z=2\times 10^{4}~\Omega$ is due to appearance of significant
radiative losses. For the considered case of even $N=10$, all
even-$n$ resonances are interfering destructively and therefore
are non-emitting. To the contrary all odd-$n$ resonances are
emitting progressively stronger with decreasing the mode number
$n$. This is most clearly seen from the voltage dependence of the
normalized emission power, shown in Fig. 1 b). It can be seen that
for $Z=10^{32} ~\Omega$ even-$n$ Fiske steps (dashed lines)
correspond to minima, while odd-$n$ steps (solid lines) to maxima
of emission. As expected, the largest (by two orders of magnitude)
emission occurs for in-phase resonances (5,1), (6,1) with the
largest Fiske step amplitudes, see Fig. 1 a). They correspond to
the velocity matching condition $m \simeq
2\mathrm{int}[\Phi/\Phi_0]$, as discussed in Ref.
\cite{SvenFiske}. The Fiske step (5,3) has larger amplitude but
smaller emission because only one third of the stack is emitting
coherently while the rest of the junctions are interfering
destructively. Also odd-$n$ Fiske steps (1,3), (3,3), (6,3),
(7,3), (4,1), (7,1) with small amplitudes, see Fig. 1 a), produce
clear emission peaks for $Z=10^{32} ~\Omega$.

Remarkably, all these emission peaks
are absent for $Z=2\times 10^{4}~\Omega$. Furthermore, the
strongest emission peaks (5,1), (6,1) at $Z=10^{32} ~\Omega$ are
replaced by minima for $Z=2\times 10^{4}~\Omega$. Apparently,
radiative losses have completely damped in-phase resonances for
$Z=2\times 10^{4}~\Omega$. This clearly illustrates that the
quality factor
is playing a central role in operation of the coherent FFO.

\begin{figure}[t]
\begin{center}
\includegraphics[width=0.8\linewidth]{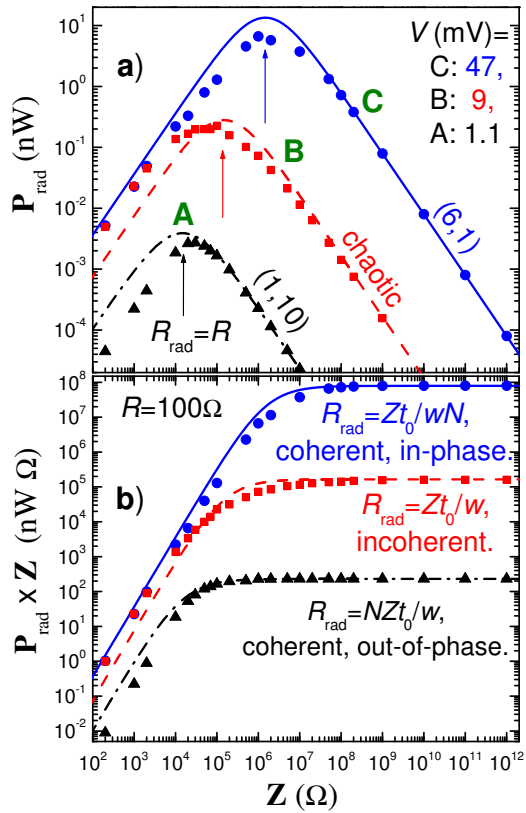}
\end{center}\label{Fig2}
\caption{(color online). Simulated emission characteristics
(symbols) as a function of the radiation impedance for three
voltages, marked in Fig. 1. A: at the out-of-phase (1,10)
resonance, B: in the incoherent state, and C: at the in-phase
(6,1) resonance. Lines represent analytic expressions Eqs.
(\ref{P_rad},\ref{PZ}) with $R_{rad}$ indicated in b). It is seen
that the emission power is decreasing both at small and large $Z$
and that the maximum power is achieved at $R_{rad}=R$, indicated
by arrows in a). Note that the same stack has different values of
$R_{rad}$, depending on collective fluxon dynamics in the stack.}
\end{figure}

The role of the output impedance on the FFO power can be
understood from the following simple model. The oscillating
amplitude at the resonance is $E_i=E_0 Q =E_0 \omega R_{eff} C$,
where $E_0$ is the amplitude out of the resonance and $R_{eff}$ is
the effective damping resistance due to both internal QP ($R$) and
radiative ($R_{rad}$) losses:
\begin{equation}\label{Reff}
R_{eff}=\left[R^{-1}+R_{rad}^{-1}\right]^{-1}.
\end{equation}
The total emission power from one side of the junction is
\begin{equation}\label{P_rad}
P_{rad}=E_0^2 \omega^2 t_0 w \frac{R^2R_{rad}}{(R+R_{rad})^2}.
\end{equation}
It is useful to consider the quantity
\begin{equation}\label{PZ}
P_{rad}Z = \gamma (1+R/R_{rad})^{-2},
\end{equation}
where $\gamma$ is constant for a given resonance $\omega_{m,n}$.

As already discussed above, see Eqs.
(\ref{Enonlocal},\ref{EmPower}), radiative losses in stacked
Josephson junctions depend essentially on the collective, rather
than local, fluxon dynamics. They can be described by the
``coherence factor":
\begin{equation}\label{CF}
C_{F}=\left|E_{ac}/E_i\right|,
\end{equation}\label{CF}
which reflects the efficiency of emission in comparison to a
single junction with the same electric field amplitude. The
effective radiative resistance (per junction) is then:
\begin{equation}\label{Rrad}
R_{rad}=R_Z/C_F.
\end{equation}

Fig. 2 shows 
a) the emission power (from one edge) and b) the product
$P_{rad}Z$, as a function of $Z$ for three voltage levels, marked
by arrows in Fig. 1: A - the coherent, phase-locked state at the
lowest out-of-phase geometrical resonance $V_{1,10}\simeq 1.1$ mV;
B - an incoherent, chaotic state at $V\simeq 9$ mV at which no
clear resonances are observed and individual junctions are having
different voltages; C - the coherent in-phase resonance
$V_{6,1}\simeq 47$ mV. Symbols and lines represent numerical
simulations and analytic expressions, Eqs. (\ref{P_rad},\ref{PZ}),
respectively. It is seen that at large $Z$, $P_{rad}Z$ is
constant. In this case radiative losses are small compared to QP
losses, $R_{rad} \gg R$, the amplitude $E_{ac}$ is independent of
$Z$ and $P_{rad}$ is increasing $\propto 1/Z$ with decreasing $Z$,
see Eq.(\ref{EmPower}). However, for $R_{rad} \lesssim R$, the
quality factor starts to decrease, resulting in reduction of
$E_{ac}$ and the emission power.

According to Eq. (\ref{P_rad}), the maximum of $P_{rad}$ is
achieved when $R_{rad}=R$. However, as follows from Eq.
(\ref{Rrad}), $R_{rad}$ for stacked junctions is not constant, but
depends on the collective state of the stack, described by the
coherence factor $C_F$. For the in-phase state C, $C_F = N$ and
$R_{rad} = R_Z/N$. The corresponding $P_{rad}$ and $P_{rad}Z$
calculated from Eqs. (\ref{P_rad},\ref{PZ}) are shown by solid
lines in Figs. 2 a) and b) and describe well the numerical data.
In this case the maximum of $P_{rad}$ is achieved at large
$Z=RNw/t_0$, marked by the blue arrow in Fig. 2 a). The
incoherent/chaotic state B, is well described by Eqs.
(\ref{P_rad},\ref{PZ}) with $N-$times larger $R_{rad}=R_Z$ (dashed
lines), because junctions in the stack act as individual single
junctions, $C_F=1$. Finally, as expected, the out-of-phase state
corresponds to the largest $R_{rad}$ which is, however not
infinite, but fairly well described by yet another $N-$times
larger $R_{rad}=NR_Z$, $C_F\simeq 1/N$, represented by the
dashed-dotted lines.

To conclude, strong emission from the stacked coherent flux-flow
oscillator requires both large coherence factor $C_F$ and large
oscillating amplitude $E_{ac}$. For a single junction FFO,
$E_{ac}$ can be greatly enhanced due to Lorentz contraction of the
fluxon \cite{Laub} at the velocity matching condition
\cite{Koshelets}. However, such mechanism is absent in stacked
Josephson junctions due to the lack of Lorentz invariance in the
system \cite{Fluxon}. Therefore, large $E_{ac}$ can be achieved
only in the presence of a high quality resonance. Importantly,
high quality geometrical resonances can impose their order on the
fluxon lattice. E.g., fast in-phase fluxon phonons may stabilize
the rectangular fluxon lattice with large $C_F=N$ even at low
fluxon velocities \cite{SvenFiske}. In addition, the resonance
reduces the linewidth of emission $\delta f/f \propto 1/Q$.
Therefore, high speed geometrical resonances are particularly
important for realization of the coherent stacked FFO. As shown
above, optimal operation of the coherent FFO is achieved when
radiative losses are equal to internal resistive losses. Finally,
it was shown that radiative losses lead to the asymmetry between
even/odd-$n$ Fiske steps in $I-V$s of stacked Josephson junctions.
This is consistent with resent experimental observations for
Bi-2212 mesas \cite{SvenFiske} and can be viewed as indirect
evidence for significant coherent flux-flow emission from
intrinsic Josephson junctions.

\end {document}